\documentclass[aps,showpacs]{revtex4}
\usepackage{color,graphicx}
\usepackage{epsfig,amsmath}
\def\roughly#1{\mathrel{\raise.3ex\hbox{$#1$\kern-.75em%
\lower1ex\hbox{$\sim$}}}}
\def\lsim{\roughly<}

\usepackage{ulem}
\usepackage{xcolor}
 
\begin{document}
\title{Phase transition towards strange matter}

\author{F. Gulminelli$^{1,2}$ }
\author{Ad. R. Raduta$^{3}$}
\author{M.Oertel$^{4}$}

\affiliation{$^{1}$~CNRS, UMR6534, LPC ,F-14050 Caen c\'edex, France\\
$^{2}$~ENSICAEN, UMR6534, LPC ,F-14050 Caen c\'edex, France\\
$^{3}$~IFIN-HH, Bucharest-Magurele, POB-MG6, Romania\\
$^{4}$~LUTH, CNRS, Observatoire de Paris, Universit\'e Paris Diderot, 5 place
Jules Janssen, 92195 Meudon, France
}

\begin{abstract}
  The phase diagram of a system constituted of neutrons and $\Lambda$-hyperons
  in thermal equilibrium is evaluated in the mean-field approximation. It is
  shown that this simple system exhibits a complex phase diagram with first
  and second order phase transitions. Due to the generic presence of
  attractive and repulsive couplings, the existence of phase transitions
  involving strangeness appears independent of the specific interaction
  model. In addition we will show under which conditions a phase transition
  towards strange matter at high density exists, which is expected to persist
  even within a complete treatment including all the different strange and
  non-strange baryon states. The impact of this transition on the composition
  of matter in the inner core of neutron stars is discussed.
\end{abstract}

\today

\maketitle

\section{Introduction}

With the purpose of better understanding the dynamics of core-collapse
supernova and the observed characteristics of neutron stars, a considerable
theoretical effort has been undertaken in recent years concerning the
modelization of the equation of state of cold dense
matter \cite{Prakash97,Lattimer06,Sedrakian06} with some extensions to finite
temperature, see
e.g. \cite{Lattimer91,Shen98,Hempel09,Blinnikov09,Heckel09,Raduta10}.

If it is well admitted that hyperonic and deconfined quark matter could exist
in the inner core of neutron stars, a complete understanding of the
composition and equation of state of dense matter is far from being achieved.
Concerning hyperons, simple energetic considerations suggest that they should
be present at high density \cite{Glendenning82}.  However, in the standard
picture the opening of hyperon degrees of freedom leads to a considerable
softening of the equation of state \cite{Glendenning82,Baldo99, Vidana00,
  Djapo10, Stone10,Schulze11, Burgio11,Massot12}, which in turns leads to
maximum neutron star masses smaller than the highest values obtained in recent
observations \cite{Demorest10}. This puzzling situation implies that the
hyperon-hyperon and hyperon-nucleon couplings must be much more repulsive at
high density than presently assumed 
\cite{Hofmann00,Bonanno11,Weissenborn11b,Weissenborn11c,Bednarek11,Lastowiecki11,Oertel12}, 
and/or that something
is missing in the present modelization.

Apart from neutron star observations, already for purely nuclear matter,
stringent constraints on the equation of state from experimental data as well
as from theoretical side from ab-initio calculations only exist up to roughly
saturation density, mainly for almost symmetric matter at zero temperature.
What makes the description of hyperonic matter even more difficult is first of
all the fact that, contrary to nucleons, hyperonic data from hypernuclei (see
e.g. \cite{Chrien90,Schaffner93,Gal10}), diffusion and production experiments
(see e.g. \cite{Saha04}) and nuclear collision experiments (see
e.g. \cite{Sagert07,Sagert11}) are scarce. This lack of experimental
information induces large uncertainties within the microscopic
approaches \cite{Baldo99,Vidana00,Djapo10}, which suffer in addition probably
from theoretical shortcomings, among others due to the unknown hyperonic
three-body forces \cite{Vidana10b}. Phenomenological extrapolations of the
low-density behavior within mean field models are subject to large
uncertainties, too. In any case, there is much uncertainty on the hyperonic
interactions at high density, where neutron star observations can give the
only hint, though not decisive.

Here, we propose to study the phase diagram of hyperonic matter. 
The generic presence of attractive and
repulsive couplings suggests the existence, in a model-independent manner, of a
phase transition involving strangeness. 
In order to have an analytically solvable model, we consider the
simplified situation where only neutrons, $n$, and $\Lambda$-hyperons
are allowed in the matter chemical composition. 
We find, as argued above, a first order phase transition involving
strangeness. In addition, we will show under which assumptions
on the $\Lambda\Lambda$ and $n\Lambda$ interaction, respecting the available
constraints, a phase transition towards strange matter, that is with 
a discontinuity in the strangeness content of matter, exists at high
density.  

We leave the
inclusion of protons and higher mass strange and non-strange baryons
to a future study. Because of this simplification, we will not be able
to exploit the results for a predictive quantitative application to
neutron stars and dense supernova matter. However, this simplification
will allow us to have an exactly solvable model with perfectly
controlled numerics.  We believe that the relevant degrees of freedom
to explore the opening of the strangeness channels are included
already at this level, and the qualitative features of the strangeness
phase transition will not change in the complete model.

Concerning the phenomenological consequences of our findings let us stress that
many published work on hyperonic matter makes use of the mean field
approximation (e.g. \cite{Glendenning82, Stone10,
  Massot12,Hofmann00,Bonanno11,Weissenborn11b, Weissenborn11c, Bednarek11,
  Lastowiecki11, Oertel12}).  However, if the opening of the strangeness
degree of freedom at high baryonic density is associated with a first order
phase transition, the mean field equations of state should be modified making
use of the Gibbs construction \cite{Glendenning01}.The possible presence of
coexisting phases would in addition modify the matter composition with respect
to the uncorrected mean-field predictions \cite{Lavagno10,Yang03,Yang05}.

Both, matter composition and equation of state, are important ingredients not
only in the calculations of neutrons star models, but in the hydrodynamical
codes describing core collapse dynamics, too. The high density and high
temperature behavior not only influences the success of the explosion, but
among others also the dependence of the final state (neutron star or black
hole) on the progenitor mass (see e.g. \cite{Nakazato11,O'Connor10}). This
means that the possible presence of a phase transition towards strange matter
has to be studied at finite temperature, too.
 
\section{The model}
In the following we will  illustrate the propositions concerning the
phase diagram by choosing a specific interaction model for the
$n\Lambda$ system. Of course, the quantitative numerical results are
not model-independent, but we will argue in which way  
our findings on the phase diagram are general.

The energetics of the $n\Lambda$ mixture is described through the
energy density functional proposed by Balberg and Gal \cite{Balberg97}
\begin{eqnarray}
\epsilon_{pot}\left(\rho_n,\rho_\Lambda\right)&=& 
\frac{1}{2}\left [\left(a_{NN}+b_{NN}\right)\rho_n^2+c_{NN}\rho_n^{\delta+1}\right]+
\frac{1}{2}\left [a_{\Lambda\Lambda}\rho_\Lambda^2
+c_{\Lambda\Lambda}\rho_\Lambda^{\gamma+1}\right] \\ \nonumber
&+&a_{\Lambda n} \rho_n \rho_{\Lambda} + c_{\Lambda n}\frac{\rho_n \rho_\Lambda}{\rho_n+\rho_\Lambda}
\left [ \rho_n^\gamma+\rho_\Lambda^\gamma\right ],
\end{eqnarray}
where the interaction couplings, compatible with the Lattimer-Swesty
equation of state in the non-strange sector, are given in table I.
Parametrizations BGI, BGII and BGIII are taken from the work
by Balberg and Gal \cite{Balberg97} and 220g2.8 corresponds to one of the
parametrizations compatible with the observation of an almost two solar
mass neutron star \cite{Demorest10} from Ref. \cite{Oertel12}. In the
applications shown below we will use mainly parametrization (BGI), but
we have performed the calculations for the other parameter sets, too.
\begin{table}
\begin{center}
\caption{ Different parameter sets for the $n-n$, $n-\Lambda$ and 
$\Lambda-\Lambda$ interaction from Ref. \cite{Balberg97} (BGI, BGII, BGIII)
and Ref. \cite{Oertel12} (220g2.8).}
\begin{tabular}{cccccccccc}
\hline
\hline
Parameter set & $a_{NN}$ & $b_{NN}$  & $c_{NN}$ &
$a_{\Lambda\Lambda}$ & $c_{\Lambda\Lambda}$ & $a_{\Lambda N}$ &
$c_{\Lambda N}$ & $\delta$ & $\gamma$ \\
 & MeV fm$^3$ & MeV fm$^3$   & MeV fm$^{3\delta}$  &  MeV fm$^3$  &
MeV fm$^{3\gamma}$  & MeV fm$^3$  &  MeV fm$^{3\gamma}$  &  & \\
\hline
(BGI)             &-784.4 &  214.2 & 1936. & -486.2 & 1553.6 & -340. &
1087.5 & 2 & 2 \\
(BGII)             &-935.4 &  214.2 & 1557.2 & -552.6 & 1055.4 & -387. &
738.8 & 5/3  & 5/3\\
(BGIII)             &-1384.6 &  214.2 & 1672.8 & -723.2 & 869.0 &
-505.2 & 605.5 & 4/3 & 4/3\\
(220g2.8) &-1636.2 &214.2 &1869.26 &-400 &1500 &-270 &2300 & 1.26 & 2.8\\
\hline
\end{tabular}
\end{center}
\label{table:balb}
\end{table}
%
\begin{figure}
\begin{center}
\includegraphics[angle=0, width=0.5\columnwidth]{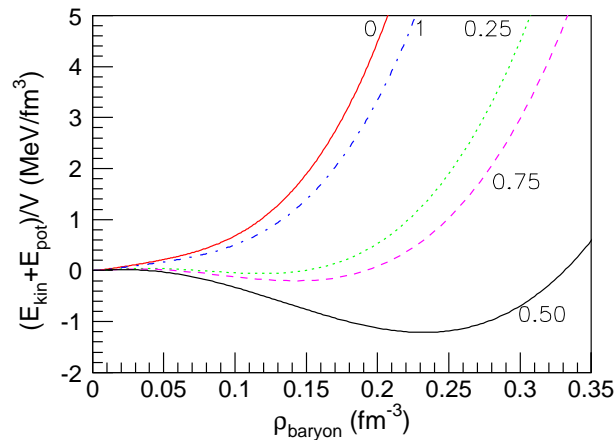}
\end{center}
\caption{(Color online) 
Energy density as a function of the total baryon
  density for a $(n,\Lambda)$ mixture at $T=0$ with different
  $\Lambda$ fractions as obtained by the parameter set (BGI).} 
\label{fig:energy_vs_rhob}
\end{figure}
In the non-relativistic mean-field approximation the kinetic energy density
$\epsilon_{kin}=\epsilon_{kin,n}+\epsilon_{kin,\Lambda}$ 
has the simple ideal Fermi gas form
\begin{equation}
\epsilon_{kin,q}=\frac{2\pi}{\beta h^3}\left (2s_q+1\right) \left (\frac{2m^*_q}{\beta}\right )^{3/2}\int_0^\infty dx\frac{x^{3/2}}{1+\exp\left (x-\beta\tilde{\mu}_q\right )},
\end{equation}
where $q=n,\Lambda$, $s_q$ is the particle spin, $\beta=T^{-1}$ is the
inverse temperature, $m^*_q=m_q$ for the chosen interaction
parameters, and the Fermi integral depends on an effective chemical
potential $\tilde{\mu}_q=\mu_q-U_q-m_qc^2$, shifted with respect to
the thermodynamic chemical potential because of the rest mass and the
depth of the self-consistent mean field $U_q\equiv \partial
\epsilon_{pot}/\partial \rho_q$. At zero temperature the Fermi
integral can be analytically solved giving
\begin{equation}
\epsilon_{kin,q}\left(T=0\right)=\frac{\hbar^2}{2m^*_q}\frac{3}{5}\rho_q
 \left (\frac{6\pi^2\rho_q}{2s_q+1}\right )^{2/3}. 
\end{equation}
The energy density $\epsilon=\epsilon_{kin}+\epsilon_{pot}$ at
zero temperature obtained with the parameter set (BGI) for different
hyperon fractions $Y_{\Lambda}=\rho_{\Lambda}/(\rho_{\Lambda}+\rho_n)$
is represented in Fig. \ref{fig:energy_vs_rhob}. 
We can see that pure neutron matter, as well as pure $\Lambda$ matter, are
never bound. For pure neutron matter this is well known. For pure
$\Lambda$-matter it is less obvious, since it involves the
$\Lambda\Lambda$-interaction, subject to large uncertainties. A strong
attraction at low densities could alter this result. We think, however, that
this is not the case, since parameter set I by Balberg and
Gal \cite{Balberg97}, shown in Fig. \ref{fig:energy_vs_rhob}, assumes a much 
stronger attraction than indicated by more recent analysis \cite{Nakazawa10}.
There is almost no doubt that the $n\Lambda$ interaction should be
attractive at low densities and repulsive at high densities. 
Due to the attractive part of the $n\Lambda$
coupling, a mixture of the two particle species admits a bound state at
finite density.  This rather general feature of
the energy density is found within the other parametrizations and other
models, e.g. the parametrization of the energy density functional from
G-matrix calculations by Vidana et al. \cite{Vidana10a}. 
Within parametrization (BGI), the lowest energy
corresponds to a symmetric mixture $\rho_n=\rho_\Lambda$ and a bound
state is predicted for $0.19<Y_{\Lambda}<0.85$. 

The existence of a minimum in the energy functional for symmetric matter
means that such a state represents the stable matter phase at zero temperature.
On the other side, the vanishing density gas phase is always the stable phase
in the limit of infinite temperature. 
This implies that a dilute-to-dense phase change
implying strangeness has to be expected on very general grounds. 
The presence of a minimum is however not sufficient to discriminate 
between a smooth cross-over and a phase transition. 
In the next section we will therefore demonstrate the existence of a 
phase transition by explicitly calculating the $n\Lambda$ phase diagram.

\section{Phase diagram of the two-component $n-\Lambda$ system}
\label{sec:phase}
First order transitions
are signaled by an instability or concavity anomaly in the mean-field
thermodynamic potential, which has to be cured by means of the Gibbs phase
equilibrium construction at the thermodynamic limit. For this reason the
convexity analysis of the thermodynamical potential in the extensive variable
space has been often employed to spot the presence of phase
transitions, e.g. for the neutron-proton system
\cite{Ducoin06,Ducoin08}. At
zero temperature, one thermodynamic
potential is given by the total energy
\begin{figure}
\begin{center}
\includegraphics[angle=0, width=0.6\columnwidth]{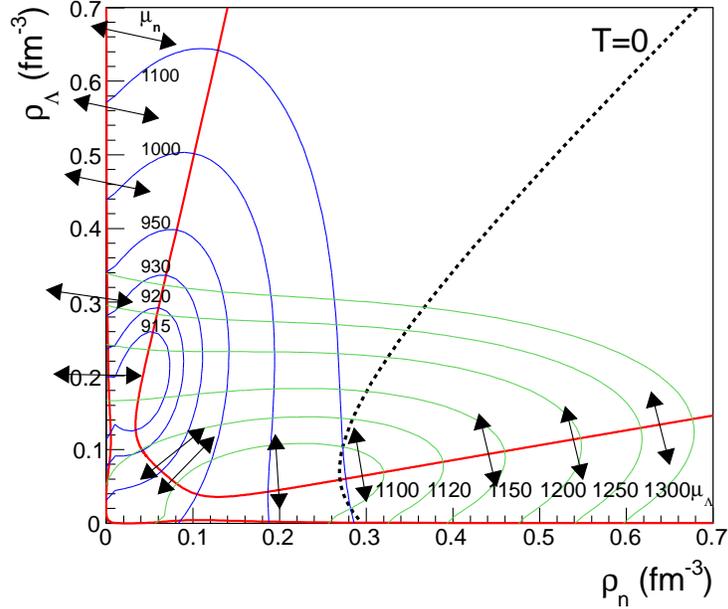}
\end{center}
\caption{(Color online)
Frontiers of the spinodal instability domain of the 
$(n,\Lambda)$ mixture at $T=0$ (thick red solid lines) corresponding 
to the parameter set (BGI);
trajectories of constant-$\mu_n$ (blue dotted lines) 
and constant-$\mu_{\Lambda}$ (green dashed lines);
the arrows indicate the directions of phase separation;
the (black) short-dashed line illustrates the $\mu_S=0$ trajectory 
associated to the equilibrium of strangeness. 
}
\label{fig:spinodal}
\end{figure}
%
\begin{figure}
\begin{center}
\includegraphics[angle=0, width=0.5\columnwidth]{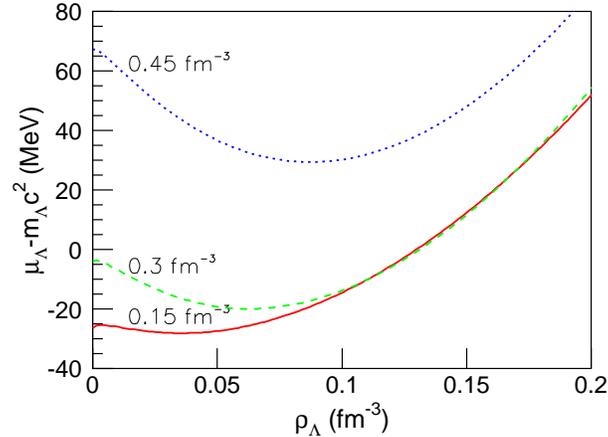}
\end{center}
\caption{(Color online)
Strangeness chemical potential as a
  function of strangeness density, $\rho_s = - \rho_{\Lambda}$, for
  different neutron densities at $T = 0$. 
The considered parameter set is (BGI).  }
\label{fig:mus_vs_rhos}
\end{figure}
%
\begin{equation}
\epsilon_{tot}\left (\rho_n,\rho_\Lambda \right)=\epsilon_{pot}\left (\rho_n,\rho_\Lambda \right)+\epsilon_{kin}\left (\rho_n,\rho_\Lambda \right)+\left (\rho_n m_n +\rho_\Lambda m_\Lambda\right ) c^2 
\end{equation}
and the curvature matrix is defined by
\begin{equation}
 \left( \begin{array}{cc}
 C_{nn} &  C_{n\Lambda} \\
C_{\Lambda n} & C_{\Lambda\Lambda}
 \end{array} \right)
= \left( \begin{array}{cc}
\frac{\partial^2 e_{tot}}{\partial \rho_n^2} & \frac{\partial^2 e_{tot}}{\partial \rho_n \partial \rho_{\Lambda}} \\
\frac{\partial^2 e_{tot}}{\partial \rho_{\Lambda} \partial \rho_n} & 
\frac{\partial^2 e_{tot}}{\partial \rho_{\Lambda}^2}
 \end{array} \right)
=\left( \begin{array}{cc}
\frac{\partial \mu_n}{\partial \rho_n} & \frac{\partial \mu_n}{\partial \rho_{\Lambda}}\\
\frac{\partial \mu_{\Lambda}}{\partial \rho_n} & 
\frac{\partial \mu_{\Lambda}}{\partial \rho_{\Lambda}}
\end{array} \right) ,
\end{equation}
with $C_{n \Lambda}=C_{\Lambda n}$ and real eigenvalues $C_{min}<C_{max}$.  
The spinodal region is then
recognized as the locus of negative curvature of the energy surface,
$C_{min}<0$. The corresponding eigenvector defines a direction in the density
space given by
\begin{equation}
\frac{\rho_n}{\rho_{\Lambda}}=\frac{C_{n \Lambda}}{C_{min}-C_{nn}}=
\frac{C_{min}-C_{\Lambda \Lambda}}{C_{\Lambda n}}.
\end{equation}
This instability direction physically represents the chemical composition of
density fluctuations which are spontaneously and exponentially amplified in
the unstable region in order to achieve phase separation, and gives the order
parameter of the associated phase transition. In all the parametrizations we
have analyzed one of the eigenvalues is always positive, meaning that the
order parameter of the $n\Lambda$ transition is always one-dimensional,
similar to the liquid-gas nuclear phase transition at sub-saturation
densities.

\begin{figure}
\begin{center}
\includegraphics[angle=0, width=0.6\columnwidth]{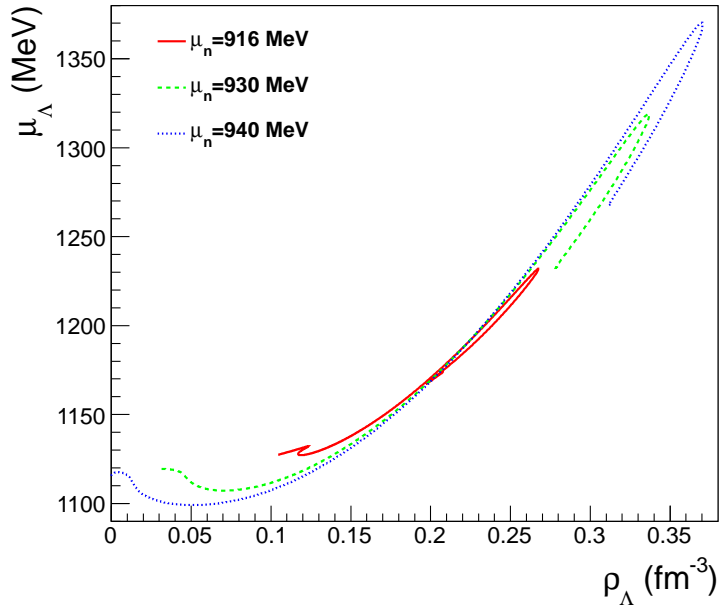}
\end{center}
\caption{(Color online)
$\Lambda$ chemical potential as a function of the associated density
$\rho_{\Lambda}$ at $T=0$ for different values of 
$\mu_n$=916, 930 and 940 MeV and (BGI). 
}
\label{fig:mul_rhol}
\end{figure}
The zero temperature instability region of the $n\Lambda$ mixture is shown in
Fig. \ref{fig:spinodal} with parameter set (BGI).
We can see that a large portion of the phase diagram is concerned by the
instability. We can qualitatively distinguish three regions characterized by
different order parameters. Below nuclear saturation density, we observe an
isoscalar $\rho_n\approx\rho_\Lambda$ instability, very close to ordinary
nuclear liquid-gas, with $\Lambda$'s playing the role of protons.  This could
have been expected, due to the similarity between the energy density for
$Y_\Lambda=0.5$ (cf Fig. \ref{fig:energy_vs_rhob}) and the well-known energy
density for low density $np$ matter.  Neutron matter close to saturation is
stabilized by the inclusion of a non-zero $\Lambda$-fraction, which is
consistent with the observation that the neutron drip-line is shifted towards
more neutron-rich systems in hypernuclei \cite{Vidana98}.

\begin{figure}
\begin{center}
\includegraphics[angle=0, width=0.4\columnwidth]{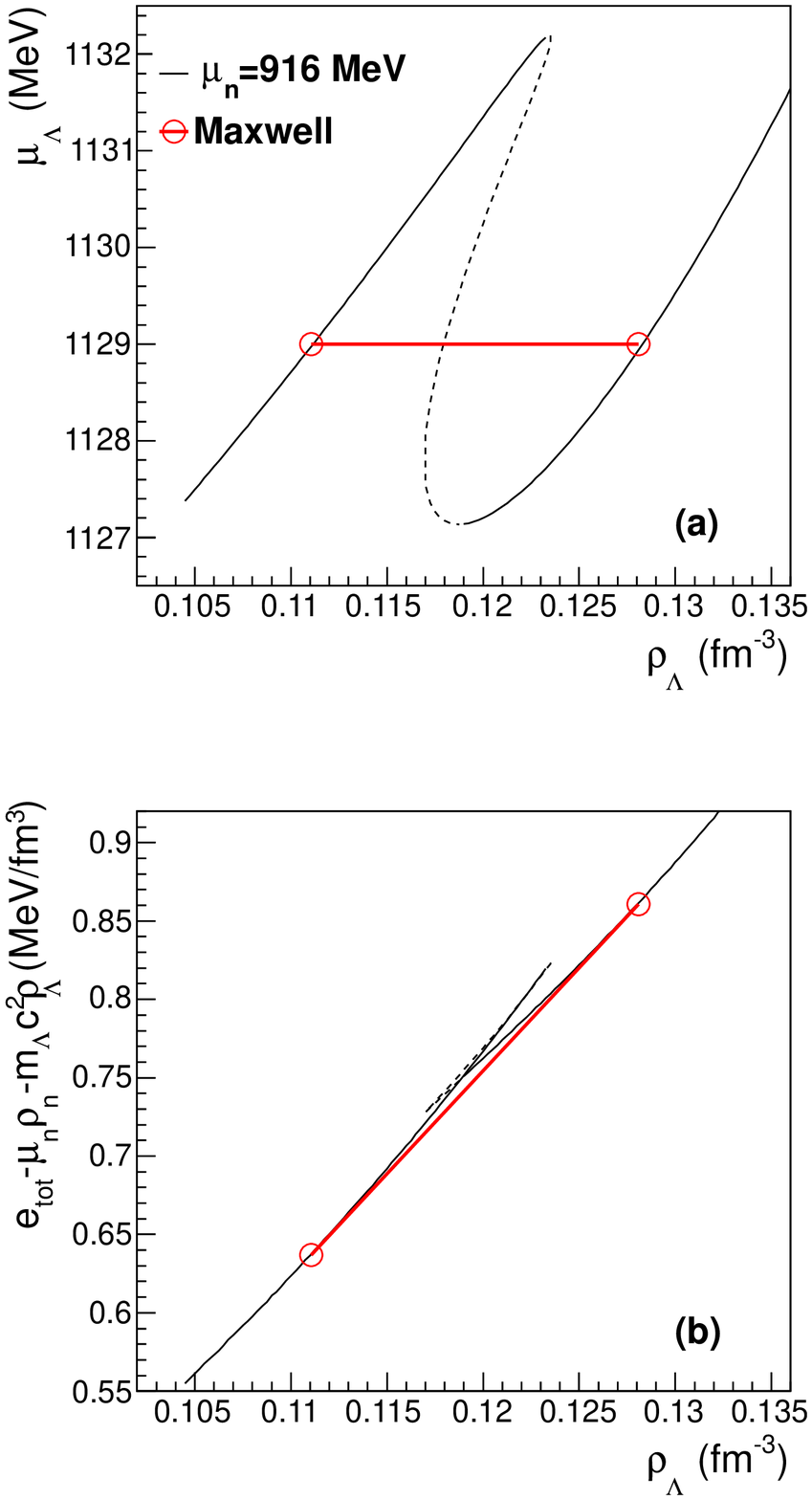}
\includegraphics[angle=0, width=0.4\columnwidth]{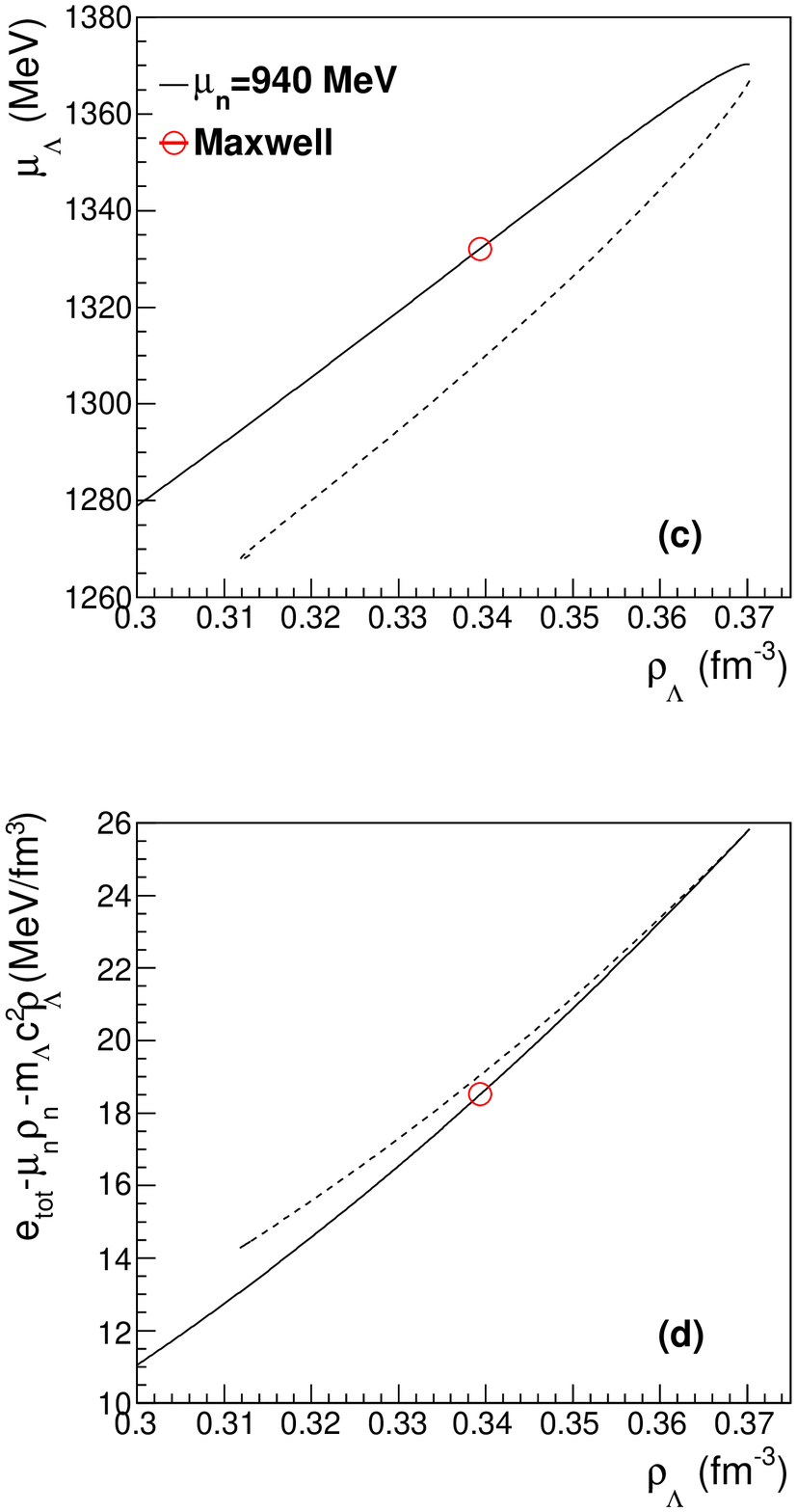}
\end{center}
\caption{(Color online)
Upper part: zoom on the low (left) and high (right) density domains
  of $\mu_{\Lambda}(\rho_{\Lambda})|_{\mu_n}$ (top) for $\mu_n=916$ MeV (left)
  and $\mu_n = 940$ MeV (right). Lower part: associated constrained energy.
  In case of multiple evaluation, the unfavorised solution(s) are represented
  with dashed line.  The red open circles and line mark the convex envelope
  defined by the Gibbs construction. 
For the density energy functional we have used (BGI). }
\label{fig:mul_rhol_zoom}
\end{figure}
Increasing neutron density the phase separation progressively changes towards
the strangeness $\rho_S=-\rho_\Lambda$ direction: the two stable phases
connected by the instability have close baryon densities but a very different
fraction of $\Lambda$'s. For high $\rho_{\Lambda}$ we observe the same
behavior with the roles of neutrons and $\Lambda$-hyperons exchanged.  The
part of the phase diagram at high neutron density comprises the region
physically explored by supernova and neutron star matter, which is
characterized by chemical equilibrium for reactions implying strangeness,
$\mu_S=\mu_n-\mu_\Lambda=0$. The strangeness equilibrium trajectory is
represented by a dashed line in Fig. \ref{fig:spinodal}.  This physically
corresponds to the sudden opening of strangeness, observed in many
modelizations of neutron star matter (see e.g. \cite{Balberg97, Massot12}).

\begin{figure}
\begin{center}
\includegraphics[angle=0, width=0.6\columnwidth]{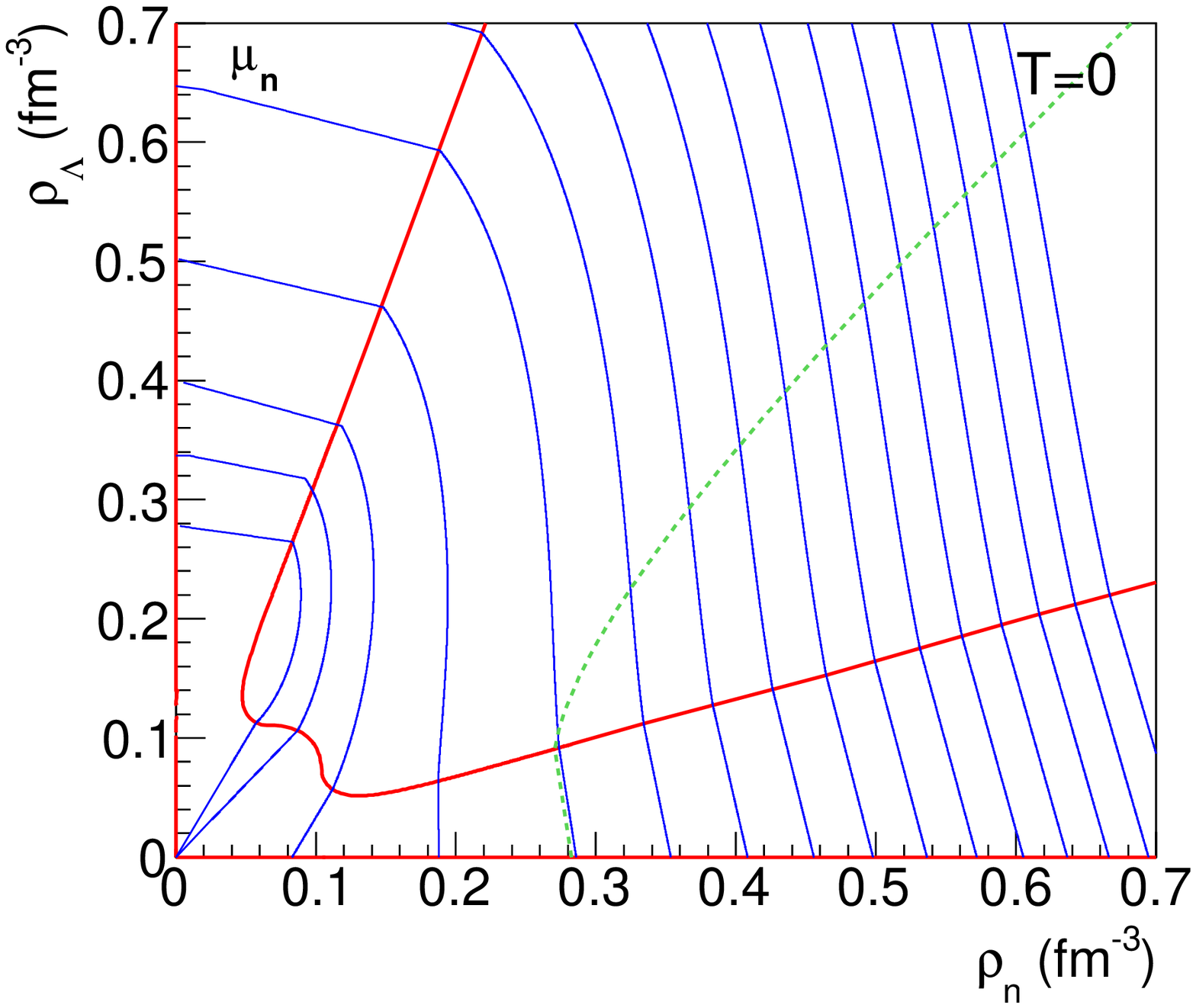}
\end{center}
\caption{(Color online)
$(n,\Lambda)$ mixture at $T=0$:
Borders of the phase coexistence region (red thick lines),
constant-$\mu_n$ paths after Maxwell construction (blue thin lines)
and $\mu_n=\mu_{\Lambda}$ trajectory after Maxwell construction 
(dark green dashed line).
>From left to right, the considered values of $\mu_n$ are
920, 930, 950, 1000, 1100, 1200, 1300, 1400, 1500, 1600, 1700, 1800,
1900, 2000, 2100 and 2200 MeV.
The considered parameter set is (BGI).
}
\label{fig:coex_contoursmun}
\end{figure}
Assuming the order parameter to be given exactly by $\rho_S$, which is a very
good approximation at high $\rho_n$, 
we can understand the existence of this high density
strangeness phase transition in terms of the $n\Lambda$ and
$\Lambda\Lambda$-interaction. Under this assumption, the curvature analysis
can be performed one-dimensionally, as a function of $\rho_{\Lambda}$ only.
Thus the instability region is determined mainly by the condition
$C_{\Lambda\Lambda} < 0$. In Fig. \ref{fig:mus_vs_rhos} the $\Lambda$ chemical
potential, with the constant mass subtracted, that is the first derivative of
the energy density with respect to $\rho_{\Lambda}$, is displayed for
parametrization (BGI). 
A minimum in $\mu_{\Lambda}$ is related to a zero in the
curvature in the strangeness direction, indicating thus the border of an
instability region. This minimum is clearly visible in
Fig. \ref{fig:mus_vs_rhos}. It is more pronounced with increasing neutron
density and shifted to higher values of $\rho_{\Lambda}$. Apart from the
trivial kinetic term $\sim \rho_{\Lambda}^{2/3}$, $\mu_{\Lambda}$ contains the
$\Lambda$ single particle potential, $U_{\Lambda}(\rho_{\Lambda},\rho_n)
= \partial \epsilon_{\mathit{pot}}/ \partial \rho_{\Lambda}$, thus reflects the
$n\Lambda$ and $\Lambda\Lambda$ interaction. Within the model by Balberg and
Gal \cite{Balberg97}, the attractive part of the $\Lambda\Lambda$ interaction
and the specific form of the $n\Lambda$ interaction contribute to this
minimum. 

\begin{figure}
\begin{center}
\includegraphics[angle=0, width=0.6\columnwidth]{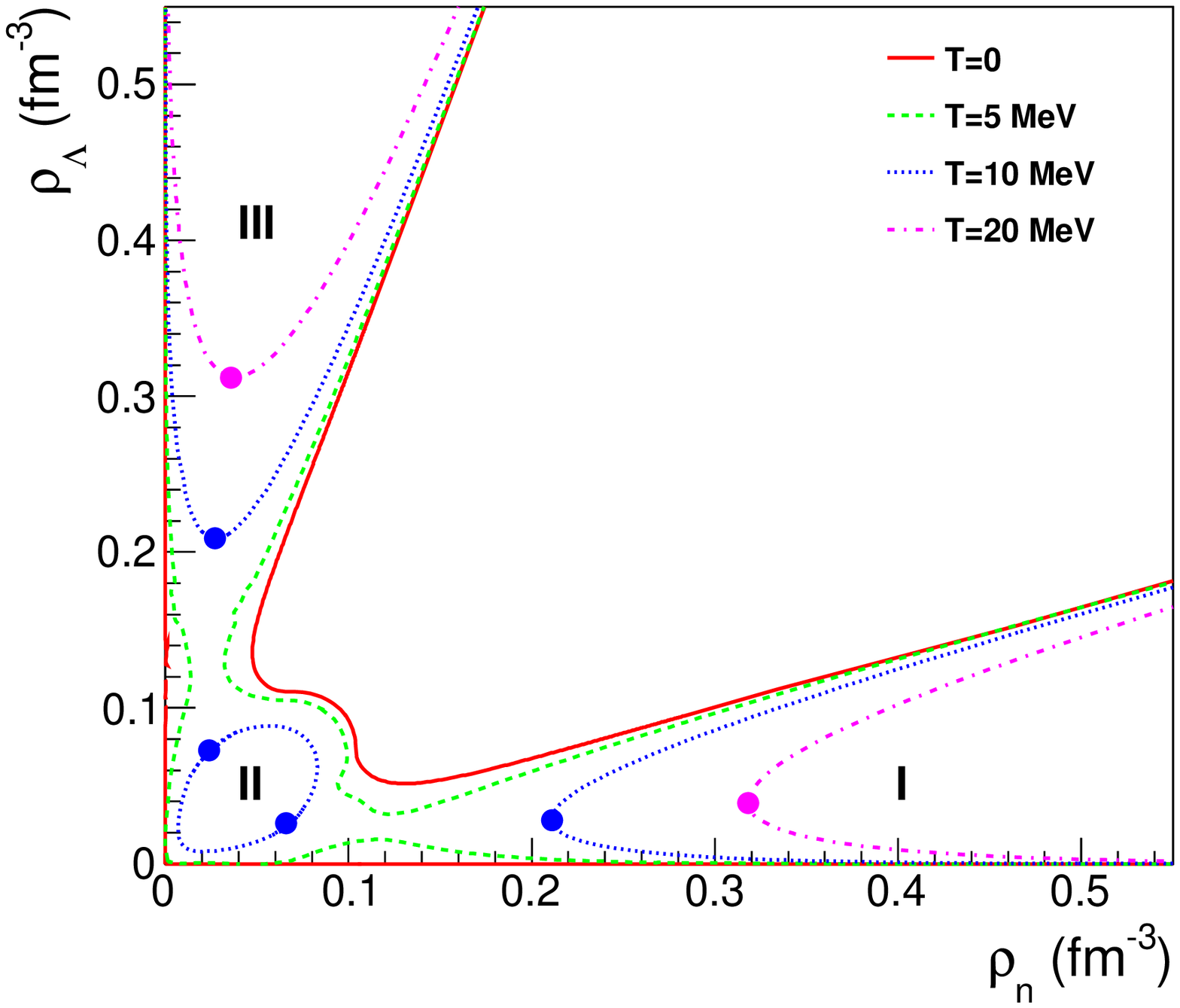}
\end{center}
\caption{(Color online)
Borders of the phase coexistence region of the 
$n\Lambda$ mixture for different temperatures, $T$=0, 5, 10, 20 MeV and (BGI).
The full circles mark the critical points. 
}
\label{fig:coex_T}
\end{figure}
Let us stress that the central region of the instability domain, below
roughly saturation density is determined mainly by the fact that pure neutron
(and $\Lambda$)-matter is unbound and there is low-density attraction in the
$n\Lambda$-channel. The finding in this region thus seems qualitatively
robust. The existence of a strangeness phase transition at high density, on
the contrary, is not a general model-independent feature, although, as
mentioned above, many models show it. There are others, for instance the
G-matrix models of Refs. \cite{Vidana10a,Burgio11}, 
which do not show an instability in
this region. This can be seen from the absence of a minimum in $\mu_{\Lambda}$
as a function of $\rho_{\Lambda}$ for constant $\rho_n$. The reason is
twofold: first, a $\Lambda\Lambda$ interaction is completely missing from these
models and the $n\Lambda$-interaction has a slightly different form than in
Balberg and Gal \cite{Balberg97}. Owing to the lack of reliable information on
the hyperonic interactions, which would discriminate between different models,
we cannot affirm the existence of the strangeness phase transition, related to
this instability, but, turning the argument around, the presence of this phase
transition in a physical system would allow to learn much about the shape of
the interaction.

One remark of caution concerning the relation of the instability domain with
a phase transition is in order here. In principle, the presence of an
instability is a pathology of mean-field approaches. It is an indication that
the lowest energy (or free-energy at finite temperature) equilibrium solution
is different from the unstable mean-field one. If the equilibrium solution
corresponds to macroscopic dishomogeneities, it can and it should be recovered
from the mean-field results making use of the Gibbs construction. In this case
the convexity in the curvature matrix reflects a physical instability towards
phase separation, and the phase diagram contains a region of phase
coexistence. However, since the mean-field equations of state are by
construction analytic infinitely differentiable functions, it is possible that
the instability is due to a multiple evaluation of densities in a given point
of the phase diagram defined by the set of associated chemical potentials,
too.  This may not be pertinent in our simple model allowing only two
different particle species, but might very well occur in a more complete model
including all hyperons and resonance states \cite{Balberg97}.  In this case the
unstable solution has to be eliminated from the equilibrium landscape, without
necessarily any phase transition.

To discriminate between the two scenarii and correct the mean-field
instabilities, one has therefore to study the phase diagram in the space of
chemical potentials.  A useful trick to spot for phase transitions with more
than one conserved charge, is to perform a Legendre transform to the
statistical ensemble where all extensive variables but one are replaced by
their conjugated intensive Lagrange parameters \cite{Ducoin05}. In this
ensemble the multidimensional Gibbs equilibrium conditions reduce to a simple
Maxwell construction.

In our simple two-dimensional case we can work equivalently within the
ensemble where the neutron density is controlled, or, alternatively, within
the ensemble where the $\Lambda$-density is controlled, corresponding to the
two constrained energies:
\begin{equation}
\bar e_{\mu_{\Lambda}}\left (\rho_n \right )=e_{tot}\left (\rho_{\Lambda},\rho_n \right )-\mu_{\Lambda} \rho_{\Lambda} \;\;  , \; \;
\bar e_{\mu_n}\left (\rho_\Lambda \right )=e_{tot} \left (\rho_{\Lambda},\rho_n \right )-\mu_n \rho_n.
\end{equation}
Let us concentrate on the second representation. The behavior of the
$\Lambda$-chemical potential,
$\mu_\Lambda\equiv\partial\bar e_{\mu_n}/\partial \rho_\Lambda$, is shown for
some selected values of the neutron chemical potential in
Fig. \ref{fig:mul_rhol}. At variance with the well-known non-strange nuclear
matter $np$ system, equilibrium can only be defined within a finite interval
of density. Indeed the ending points of the curves correspond to vanishing
neutron density, as it can be seen from the iso-$\mu$ contours in
Fig. \ref{fig:spinodal} above.

\begin{figure}
\begin{center}
\includegraphics[angle=0, width=0.45\columnwidth]{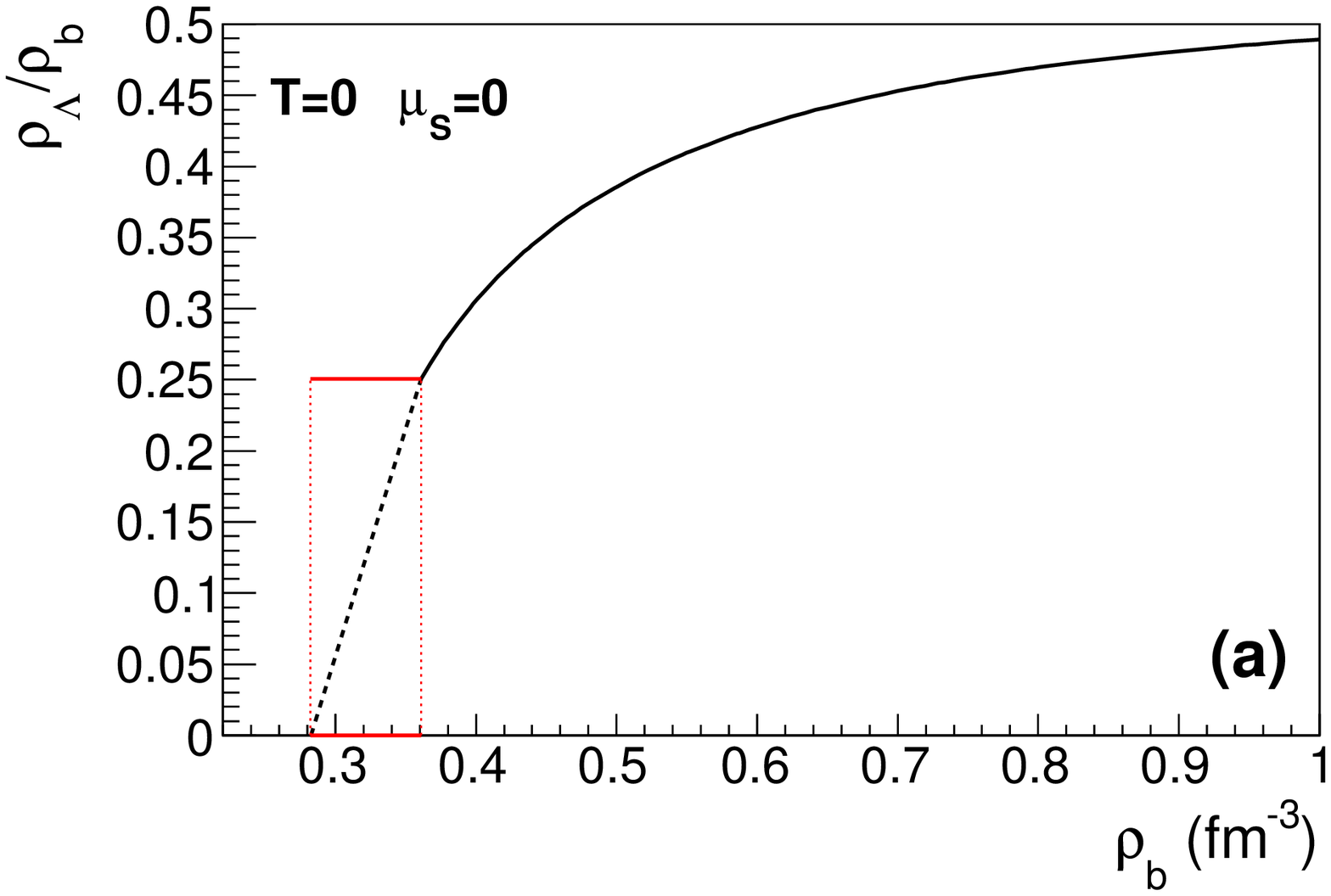}
\includegraphics[angle=0, width=0.45\columnwidth]{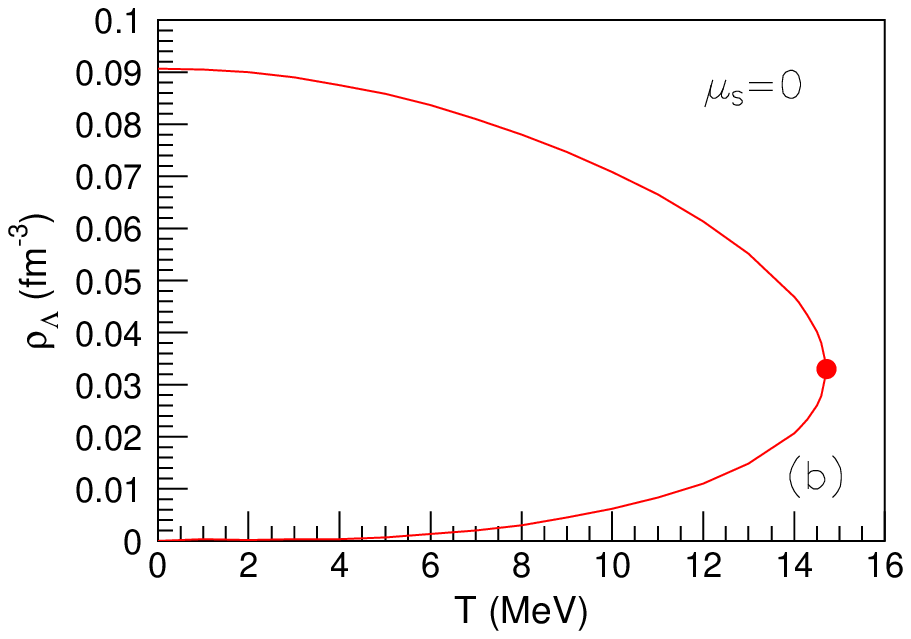}
\end{center}
\caption{ (Color online)
Left: $\Lambda$ fraction as a function of the baryon density within 
the
  condition $\mu_n=\mu_\Lambda$ for a $n\Lambda$ mixture at $T$=0.
  The full line corresponds to the stable mean-field results; 
  the dashed line illustrates the Gibbs construction; 
  the horizontal lines indicate the relative amount of $\Lambda$-hyperons 
  at the frontiers 
  of the phase-coexistence domain.
  Right: $\Lambda$ density for the two coexisting phases as a function of 
  temperature for $\mu_S=0$. 
  As before, we have used (BGI).
}
\label{fig:lambdas}
\end{figure}
At relatively high neutron chemical potential the curves present a back-bending
at low $\Lambda$ densities, similar to the usual Van der Waals phenomenology
for the fluid transition at finite temperature. In this case the mean-field
solution is unique, but a more stable solution is obtained by phase mixing. At
low $\mu_n$ values and, independent of $\mu_n$, in the high density region,
multiple evaluations are observed, where different values of $\mu_\Lambda$ are
compatible with the same controlled value of the hyperon density. In this
situation, only the solution leading to the lowest constrained energy has to
be retained.  To explore these different possibilities, a zoom of
Fig. \ref{fig:mul_rhol} for $\mu_n=916$ MeV is shown in
Fig. \ref{fig:mul_rhol_zoom}, together with the associated thermodynamic
potential. For a better visibility, the rest mass contribution $m_{\Lambda}
c^2\rho_{\Lambda}$has been subtracted, which does not alter the convexity
properties of the function.
We can see that at low density, after elimination of the least stable
solution, the constrained energy can still be minimized by taking a linear
combination of two homogeneous solutions (circles in
Fig. \ref{fig:mul_rhol_zoom}) having the same first-order derivative, that is
the same chemical potential. Since these two points have the same value for
all the intensive variables ($\mu_n,\mu_\Lambda,T$), they respect Gibbs
equilibrium rules. One can therefore see that the ensemble of Maxwell
constructions in the $(\mu_n,\rho_\Lambda)$ ensemble is equivalent to the
construction of the global convex envelope of the energy, that is to the Gibbs
construction of the complete system.

In the high density case (right part of Fig. \ref{fig:mul_rhol_zoom}), the
elimination of the multiple evaluation leaves a monotonous equation of state,
which does not allow any energy gain by linear interpolations. Going back
again to the iso-$\mu$ curves of Fig. \ref{fig:spinodal} , we can see that the
high $\rho_\Lambda$ region corresponding to the bi-evaluation can also be
explored in the complementary ($\mu_\Lambda,\rho_n)$ ensemble where it will be
associated to low values of $\rho_n$. This means that from a practical point
of view, once the phase mixing is systematically performed on the low density
($\rho_\Lambda$ and $\rho_n$) region, all the unstable mean-field solutions
turn out to be automatically removed.

The final result of the Gibbs construction is given in
Fig. \ref{fig:coex_contoursmun} together with the corrected iso-$\mu_n$ paths.
We can see that the whole unstable region of Fig. \ref{fig:spinodal} can be
interpreted as the spinodal region of a first order phase transition.  In
particular if we follow the physical path $\mu_n=\mu_\Lambda$ associated to
equilibrium of strangeness (dashed line in Fig. \ref{fig:coex_contoursmun}),
the emergence of strangeness with the opening of the hyperon channel
corresponds to the mixed-phase region of a first order phase transition, which
is not correctly modeled in the mean-field approximation.

Up to now we have only presented results at zero temperature. The extension to
finite temperature, as it is needed for supernova matter, is relatively simple
in the mean-field approximation.  The appropriate thermodynamic potential at
non-zero temperature is given by the Helmholtz free energy
\begin{equation}
f_T\left ( \rho_n,\rho_\Lambda\right)= \epsilon_{tot}\left ( 
\rho_n,\rho_\Lambda\right)-Ts\left ( \rho_n,\rho_\Lambda\right),
\end{equation}
where $s$ is the mean-field entropy density.  Chemical potentials can be
obtained by differentiating this expression or, in a simpler numerical way, by
inverting the Fermi integral associated to the densities \cite{Antia93} from
the two coupled equations:
\begin{equation}
\rho_{q}=\frac{2\pi}{h^3}\left (2s_q+1\right) \left (\frac{2m^*_q}{\beta}\right )^{3/2}\int_0^\infty dx\frac{x^{1/2}}{1+\exp\left (x-\beta\tilde{\mu}_q\right )},
\end{equation}
with $q=n,\Lambda$.  Then the Gibbs construction is performed, as in the zero
temperature case, from the combined analysis of $\mu_n(\rho_n)$ at constant
$\mu_\Lambda$, and $\mu_\Lambda(\rho_\Lambda)$ at constant $\mu_n$.  The same
qualitative behaviors as in Fig. \ref{fig:mul_rhol} are observed, with the
difference that at finite temperature chemical potentials tend to $-\infty$
with vanishing density. As a consequence, the lower energy border of the
coexistence zone is always at finite non-zero density, and Gibbs constructions
are always equivalent to equal-area constructions in the mono-extensive
ensemble.

The phase diagram as a function of the temperature is presented in
Fig. \ref{fig:coex_T}.  This phase diagram exhibits different interesting
features.  We can see that the three regions that we have tentatively defined
at zero temperature appear as distinct phase transitions at finite
temperature. The first phase transition (zone II in Fig. \ref{fig:coex_T})
separates a low density gas phase from a high density more symmetric liquid
phase, very similar to ordinary liquid-gas.  The second one (zone III in
Fig. \ref{fig:coex_T}) reflects the instability of dense strange matter towards
the appearance of neutrons and has an almost symmetric counterpart (zone I
in Fig. \ref{fig:coex_T}) in the instability of dense neutron matter towards
the formation of $\Lambda$-hyperons. 
Up to a certain temperature, this latter phase transition is explored
by the $\mu_n=\mu_\Lambda$ trajectory, meaning that it is expected to occur in
neutron stars and supernova matter. At variance with other known phase
transitions in nuclear matter, this transition exists at any temperature and
is not limited in density; it is always associated (except at $T\lsim 5$ MeV
in the present model)
with a critical point, which moves towards high density as the temperature
increases. This means that criticality should be observed in hot supernova
matter, at a temperature which is estimated as $T_c=14.8$ MeV in the present
schematic model.

The consequences of these findings for the composition of neutron star matter
are drawn in Fig.
\ref{fig:lambdas}. The left panel shows the $\Lambda$ fraction
$Y_\Lambda=\rho_\Lambda/(\rho_n+\rho_\Lambda)$ as a function of the baryon
density under the condition $\mu_n=\mu_\Lambda$. The crossing of the
mixed-phase region with increasing neutron density implies that, as soon as
the lower density transition border is crossed, the system has to be viewed as
a dishomogeneous mixture of macroscopic regions composed essentially of
neutrons, with other macroscopic regions with around $25\%$ of hyperons. The
extension of the $\Lambda$-rich zone increases with density until the system
exits the coexistence zone, and becomes homogeneous again.


On the right panel of Fig. \ref{fig:lambdas} the $\Lambda$-densities for the
two coexisting phases are displayed as a function of temperature, again under
the condition of $\mu_n = \mu_\Lambda$, physically relevant for neutron star
and supernova matter. The extension of the coexistence region decreases with
increasing temperature. The latter finally disappears at $T_c = 14.8$
MeV. This well illustrates the fact already observed in connection with the
phase diagram, cf Fig. \ref{fig:coex_T}, that the critical point of the
strangeness phase transition moves to higher density, crossing the physical
line $\mu_n = \mu_\Lambda$ at a given temperature, $T_c = 14.8$ MeV in the
present example.  

\section{Conclusions}
In this paper we have calculated the phase diagram of an interacting system of
neutrons and $\Lambda$-hyperons in the mean field approximation. We have shown
that this simple system presents a complex phase diagram with first and second
order phase transitions. Some of these phase transitions are probably never
explored in physical systems. However, a possible phase transition at
super-saturation baryon densities, from non-strange to strange matter is
expected to be observed both in the inner core of neutron star and in the
dense regions of core-collapse supernova. For this latter phenomenology, a
critical point is predicted and the associated critical opalescence could have
an impact on supernova dynamics \cite{Margueron04}.  The existence of this
particular phase transition can be related to the form of the $\Lambda\Lambda$
and $n\Lambda$ interaction and is, in the present model, essentially due to
the presence of an attractive low density $\Lambda\Lambda$ interaction as well
as to the high-density part of the $n\Lambda$-interaction. As such, it is
expected to persist in a realistic model of dense matter including more
hyperonic and non-hyperonic baryons.  The immediate consequence of that is
that the opening of hyperon channels at high density should not be viewed as a
continuous (though abrupt) increase of strangeness in the matter, observed in
many models of hyperonic matter, cf e.g. \cite{Massot12}, but rather as the
coexistence of hyperon-poor and hyperon-rich macroscopic domains.

Different steps have however to be achieved before quantitative predictions on
neutron star physics can be drawn from this simple model. The model
should be extended to include all possible hyperons and resonances, which
could shift the coexistence borders and induce new phenomena in the direction
- not explored in this preliminary study - of the electric charge
density. These improvements will be the object of a future publication.
 
\begin{acknowledgments}
{This paper has been partly supported by ANR under the project NEXEN 
and the project SN2NS, ANR-10-BLAN-0503, and by IFIN-IN2P3 agreement
nr. 07-44.
Support from Compstar, a research networking programme of the European Science
foundation, is acknowledged, too.
Ad. R. R acknowledges partial support from the Romanian National Authority 
for Scientific
Research under grant PN-II-ID-PCE-2011-3-0092 and kind
hospitality from LPC-Caen.}
\end{acknowledgments}


\end{document}